\documentstyle[aps,amsfonts,amssymb,multicol,prl]{revtex}
\input{psfig.sty}
\begin{document}
\draft
\def\vareps{\varepsilon}
\title{The Kondo Box: A Magnetic Impurity in an Ultrasmall Metallic Grain}
\author{Wolfgang B. Thimm$^{1}$, Johann Kroha$^{2}$, and Jan von Delft$^{1}$}
\address
{$^{1}$Institut f\"ur Theoretische Festk\"orperphysik,
 $^{2}$Institut f\"ur Theorie der Kondensierten Materie,\newline
Universit\"at Karlsruhe, Postfach 6980, 76128 Karlsruhe, Germany}
\date{Submitted to Phys.\ Rev.\ Lett.\ on September 30, 1998} 
\maketitle
\begin{abstract}
We study the Kondo effect generated by a single magnetic
impurity embedded in an ultrasmall metallic grain,  
to be called a ``Kondo box''. We find that the Kondo resonance
is strongly affected when the
mean level spacing in the grain becomes larger than the Kondo
temperature, in a way that depends on
the parity of the number of electrons on the grain. 
We show that the single-electron tunneling conductance through 
such a grain features Kondo-induced
 Fano-type resonances of measurable size, 
with an anomalous dependence on  temperature and level spacing.
\end{abstract}
\begin{multicols}{2}
What happens to the Kondo effect when a metal sample  containing
magnetic impurities is made so small that its
conduction electron spectrum 
becomes discrete with a non-zero mean level spacing
$\Delta$? More specifically, when will the Kondo resonance at the Fermi
energy $\varepsilon_F$ that characterizes bulk Kondo physics
begin to be affected? 
This will occur on a scale $\Delta \simeq T_K$, the bulk Kondo temperature, 
since a fully-developed resonance
requires a finite  density of states (DOS) near $\varepsilon_F$,
 and $\Delta$ will act as low-energy
cut-off for the spin scattering amplitude.

To achieve $\Delta \gtrsim T_K$, the sample would have to be an
ultrasmall metallic grain containing magnetic impurities, to be called
a ``Kondo box'': for example, for a metallic grain of volume
(15nm)${}^3$--(3nm)${}^3$ and $k_F \simeq 1\mbox{\AA}^{-1}$, the
free-electron estimate $\Delta = 1/N_0 \simeq 2 \pi^2 \hbar^2 / (m k_F
\! \mbox{{\it Vol}\/})$, with $ N_0$ the bulk DOS near
$\varepsilon_F$, gives $\Delta \simeq 0.5 - 60$K, which sweeps a range
including many typical Kondo temperatures.  The discrete DOS of an
{\em individual}\/ grain of this size can be measured directly using
single-electron tunneling (SET) spectroscopy
\cite{ralph.97,Schmidt-97}, as shown by Ralph, Black and Tink\-ham
\cite{ralph.97} in their studies of how a large level spacing affects
superconductivity.  Analogous experiments on a Kondo box should be
able to probe how a large $\Delta (\simeq T_K)$ affects Kondo physics.

In this Letter we study this question theoretically.  We find (1) that
the Kondo resonance splits up into a series of sub-peaks corresponding
to the discrete box levels; (2) that its signature in the SET
conductance through the grain consists of Fano--like line shapes with
an anomalous  temperature dependence, estimated to be of
measurable size; (3) an even/odd effect: if the total number of
electrons on the grain (i.e.\ delocalized conduction electrons plus
one localized impurity electron) is odd, the weight of the Kondo
resonance decreases more strongly with increasing $\Delta$ and $T$
than if it is even.

{\em The model:---}\/ For the impurity concentrations of 0.01\% to
0.001\% that yield a detectable Kondo effect in bulk alloys, an
ultrasmall grain of typically $10^4-10^5$ atoms will contain only a
single impurity, so that inter-impurity interactions need not be
considered.  We thus begin by studying the local dynamics of a single
impurity in an isolated Kondo box, for which we adopt the (infinite
$U$) Anderson model with a discrete conduction spectrum, in the
slave-boson representation:
\begin{eqnarray}
H= H_0  + \varepsilon _d \sum _{\sigma } f^{\dag}_{\sigma }
f^{\phantom{\dag}}_{\sigma } + v \sum _{j,\sigma} 
(c^{\dag}_{j\sigma}b^{\dag } f^{\phantom{\dag}}_{\sigma} + h.c.) ,
\label{H}
\end{eqnarray}
where $H_0 = \sum _{j,\sigma } \vareps _j
c_{j\sigma}^{\dag}c^{\phantom{\dag}}_{j\sigma}$. 
Here  $\sigma$ denotes spin and the  
$c^\dagger_{j \sigma}$ create conduction electrons
in  the discrete,  delocalized eigenstates 
$|j \sigma \rangle$ of the ``free'' system (i.e.\ without impurity).
 Their energies, 
measured relative to the chemical potential $\mu$, 
are taken uniformly spaced for simplicity:
$\varepsilon_j = j \Delta + \bar \varepsilon_0 - \mu$.
As in \cite{vdelft.97}, we follow the so-called orthodox model
and assume that the $\varepsilon_j$'s  include all effects of 
Coulomb interactions involving delocalized electrons,
up to an overall constant, the charging energy $E_C$.
The localized level
of the magnetic impurity has bare energy $\varepsilon_d$ 
far below $\varepsilon_F$, and is represented
in terms of auxiliary fermion and boson
operators as $d_{\sigma}^{\dag} = f_{\sigma}^{\dag} b$, 
supplemented by the
constraint $\sum_{\sigma} f_{\sigma}^{\dag}f_{\sigma}+
b^{\dag}b = 1$ \cite{barnes.76},
which implements the limit $U \to \infty$ for the 
Coulomb repulsion $U$ between  two electrons on the $d$-level.
Its hybridization  matrix element $v$ with 
the conduction band is 
an overlap integral between a localized and a delocalized
wave-function, and, 
due to the normalization of the latter,
scales as $(\mbox{\it Vol\/})^{- 1/2}$. 
Thus the effective width of the $d$-level,
$\Gamma = \pi v^2 /\Delta$, is volume-independent,
as is  the  bulk Kondo temperature,
$T_K = \sqrt{2\Gamma D/ \pi  }\ 
{\rm exp}(-\pi\varepsilon _d/2\Gamma)$,
where $D$ is a high energy band cutoff.
To distinguish, within the grand canonical 
formalism,  grains for which the 
total number of electrons is  
even or odd, we choose $\mu$ either on 
($\mu = \bar \varepsilon_0$) or half-way between two 
($\mu = \bar \varepsilon_0 + \Delta/2$) single-particle
levels, respectively \cite{vdelft.97}.

{\em NCA approach:---}\/
We  calculated the spectral density $A_{d\sigma}(\omega)$ of 
the impurity Green's function $G_{d\sigma} (t) = -i \theta(t)
 \langle \{ d_\sigma (t), d_\sigma^\dagger (0)\} \rangle$ using 
the noncrossing approximation (NCA) \cite{NCA.ref}.
For a continuous conduction band, the NCA is known to be reliable
down to energies of $0.1 T_K$ or less, 
producing  spurious
singularities only for
$T$ below 
\begin{figure}
\centerline{\psfig{figure=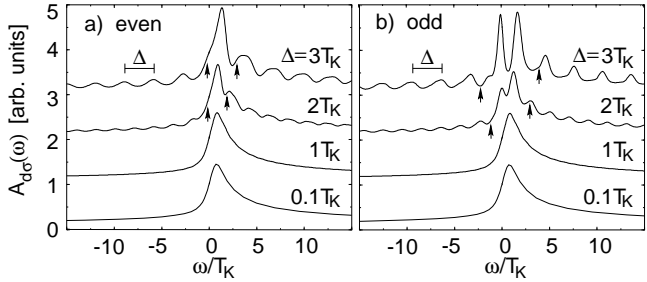,width=8.5cm}}
\narrowtext
\caption{
Impurity spectral function $A_{d\sigma}(\omega )$ for various values
of $\Delta $ at $T=0.5T_K$; (a) even, (b) odd total number of 
electrons. The individual curves are offset by one unit each.
}
\label{spectra}
\end{figure}
\noindent 
this scale \cite{bickers.87,costi.96}. Since these are cut
off by the level  spacing  
$\Delta$ in the present case, we expect 
the NCA to be semi-quantitatively accurate over 
the entire parameter range 
studied here ($T$ and $\Delta$ between 0.1 and 5$T_K$).
Denoting the retarded auxiliary fermion and boson propagators by $G_{f\sigma
  }(\omega ) =(\omega -\varepsilon _d - \Sigma _{f\sigma}(\omega ))^{-1}$,
$G_{b}(\omega ) =(\omega - \Sigma _{b}(\omega ))^{-1}$, 
respectively, the
selfconsistent NCA equations read
\begin{eqnarray}
\Sigma _{f\sigma}(\omega ) &=& \Gamma  \int \frac{d\varepsilon}{\pi}
 [1- f(\varepsilon )]  A^{(0)}_{c\sigma}(\varepsilon )
G_b(\omega - \varepsilon) \; ,  \label{NCAf}\\
\Sigma _{b}(\omega ) &=& \Gamma \sum _{\sigma}  \int \frac{d\varepsilon}{\pi}
 f(\varepsilon ) A^{(0)}_{c\sigma}(\varepsilon )
G_{f\sigma}(\omega+ \varepsilon) \, , 
\label{NCAb}
\end{eqnarray}
where $f(\omega )=1/[{\rm exp}(\omega/T )+ 1]$. 
The finite grain size enters through the discreteness of the (dimensionless)
single-particle spectral density of the box {\it without} impurity,
$A^{(0)}_{c\sigma }(\omega) = \Delta \sum _j \delta (\omega - \epsilon _j )$.
(We checked that all our results are essentially unchanged
if the Dirac $\delta$'s are slightly broadened by a width
$\gamma \lesssim 0.1 T_K$.)
In terms of the auxiliary particle
spectral functions  $A_{f,b} = - \frac{1}{\pi} {\rm Im} \, G_{f,b}$,
$A _{d\sigma}(\omega )$ is given  by
(for details see \cite{costi.96})  
\begin{eqnarray}
A _{d\sigma}(\omega ) =  \int {d\varepsilon}
[{\rm e}^{-\beta\varepsilon }+
 {\rm e}^{-\beta(\varepsilon - \omega)}] A_{f\sigma}(\varepsilon )
                             A_b(\varepsilon -\omega) \; .
\label{Ad}
\end{eqnarray}

\begin{figure}
\centerline{\psfig{figure=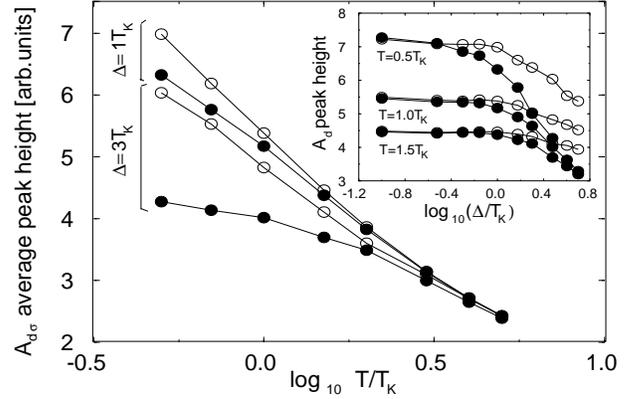,width=8cm}}
\narrowtext
\caption{
Even/odd dependence of the average peak height of the Kondo resonance,
as function of $T/T_K$. 
For an even box ($\circ$), we averaged  $A_{d\sigma}$ over  a
range  $\Delta$ centered on its central sub-peak,
for an odd box ($\bullet$) over 
a range  $2 \Delta$ centered on its central two sub--peaks 
(as indicated by arrows in Fig.~1). 
The inset shows the same quantity as  function of $\Delta/T_K$. 
Numerical uncertainties are smaller than  the symbol sizes. }
\label{sweight}
\vspace*{-2mm}
\end{figure}

{\em Numerical results:---}\/ The results obtained for $A_{d \sigma}
(\omega)$ by numerically solving the NCA equations (\ref{NCAf}) to
(\ref{Ad}) for various $T$ and $\Delta$ are summarized in
Figs.~\ref{spectra} and \ref{sweight}.  (We have checked that the
equation-of-motion method \cite{Lacroix} yields qualitatively similar
results for all quantities discussed below.) For $\Delta \ll T$, the
shape of the Kondo resonance is indistinguishable from the bulk case
($\Delta \to 0$); when $\Delta$ is increased well beyond $T$, however,
it splits up into a set of individual sub-peaks.  With decreasing
temperature (at fixed $\Delta$), each sub-peak becomes higher and
narrower; its width was found to decrease without saturation down to
the lowest temperatures for which our numerics were stable ($T\simeq
0.2 \Delta$). This agrees with the expectation following from the
Lehmann representation at $T=0$, $A_{d\sigma} (\omega) $ $ = \sum_n
\left[ |\langle n | d_\sigma^\dagger | 0 \rangle|^2 \delta (\omega -
  \tilde\Delta_n) + |\langle n | d_\sigma | 0 \rangle|^2 \delta
  (\omega + \tilde\Delta_n) \right],
$
namely that the sub-peaks should reduce to $\delta$-functions
with zero width, located at
the exact excitation energies 
$\tilde\Delta_n = E_n - E_0$ 
of the full Hamiltonian
(whose spectrum will be discrete too, with mean level spacing
of the same order as $\Delta$, as follows from the exact
finite-size solutions of the Kondo model \cite{AL91}).

Despite developing sub-peaks, the Kondo resonance retains its main
feature, namely significant spectral weight within a few $T_K$ around
the Fermi energy, up to the largest ratios of
$\Delta/\mbox{max}(T,T_K)$ ($\simeq 5$) we considered.  This implies
that the Kondo correlations induced by the spin-flip transistions
between the $d$-level and the lowest-lying unoccupied $j$-levels
persist up to remarkably large values of $\Delta/\mbox{max}(T,T_K)$
\cite{Fulde}.  However, they do weaken systematically with increasing
$\Delta$, as can be seen in the inset of Fig.~2, which shows the
average peak height of the Kondo resonance (which quantifies the
``strength'' of the Kondo correlations) as function of $\Delta$ at
fixed $T$: the peak height drops logarithmically with increasing
$\Delta$ once $\Delta$ becomes larger than about $T$.  Conversely, at
fixed $\Delta$, it drops logarithmically with increasing $T$ once $T$
becomes larger than about $0.5 \Delta$ (main part of Fig.~2), thus
reproducing the familiar bulk beha\-vior.  Qualitatively, these
features are readily understood in perturbation theory, where the
logarithmic divergence of the spin flip amplitude, $t(\omega ) \propto
\sum_{ j \neq \omega} {f(\varepsilon_j) \over \omega -
  \varepsilon_{j}}$, is cut off by either $T$ or $\Delta$, whichever
is largest.

{\em Parity Effects:---}\/ For $\Delta \gg T$, the even and odd
spectral functions $A_{d \sigma}$ in Fig.~1 differ strikingly: the
former has a single central main peak, whereas the latter has two main
peaks of roughly equal weight.  This can be understood as follows: For
an even grain, spin-flip transitions lower the energy by roughly $T_K$
by binding the $d$ electron and the conduction electrons into a Kondo
singlet, in which the topmost, {\em singly}\/-occupied $j$ level of
the free Fermi sea carries the dominant weight, hence the single
dominant peak in $A_{d\sigma}$.  For an odd grain, in contrast, the
free Fermi sea's topmost $j$ level is {\em doubly}\/ occupied,
blocking such energy-lowering spin-flip transitions.  To allow the
latter to occur, these topmost two electrons are redistributed with
roughly equal weights between this and the next-higher-lying level,
causing two main peaks in $A_{d\sigma}$ and reducing the net energy
gain from $T_K$ by an amount of order $\Delta$.  This energy penalty
intrinsically weakens Kondo correlations in odd relative to even
grains; indeed, the average $A_{d \sigma}$ peak heights in Fig.~2 are
systematically lower in odd than in even grains, and more so the
larger $\Delta$ and the smaller $T$.
 
{\em SET conductance:---}\/
The above physics should show up in SET spectroscopy 
experiments:  When an ultrasmall grain is
connected via tunnel junctions to left $(L)$ and right $(R)$
leads \cite{Schoen} and if the tunneling current through the grain
is sufficiently small (so that it only probes but
does not disturb the physics on the grain), 
the tunneling conductance $G(V)$ as function of the 
transport voltage $V$  has been demonstrated
\cite{ralph.97} to reflect the grain's discrete, equilibrium 
conduction electron DOS. 
Such measurements are parity-sensitive  \cite{ralph.97}
even though a non-zero current requires parity-fluctuations, 
since  these  can be 
minimized by exploiting the huge charging
energies ($E_c > 50$K) of the ultrasmall grain. 
To calculate the SET current, we describe 
tunneling between grain and leads by
$H_t = \sum _{kj\sigma\alpha} (u_{kj\sigma}^{\alpha } 
c^\dagger_{k\sigma\alpha} c_{j\sigma}  +h.c.)$, 
where $c^\dagger_{k\sigma\alpha}$
creates a spin $\sigma$ electron in  channel $k$ of lead 
$\alpha \in \{L,R\}$. 
Neglecting non-equilibrium effects in the grain, 
the tunneling current has the 
Landauer--B\"uttiker form \cite{wingreen.92}
\begin{equation}
\label{eq:Landauer}
  I(V) = {e \over \hbar} \int \! d \omega \, F_V (\omega)
\sum_{j\sigma} 
\left[{\gamma^L \gamma^R \over \gamma^L + \gamma^R} 
\right]_{j\sigma} A_{c, j\sigma} (\omega)  , 
\end{equation}
where $F_V(\omega ) = f(\omega -eV/2)-f(\omega +eV/2)$,
$A_{c, j\sigma}$ is the spectral density of 
$G _{c, j\sigma} (t) = 
-i \theta(t) \langle \{ c_{j \sigma}(t), c^\dag_{j\sigma} (0) \} \rangle$,
and $\gamma^\alpha_{j\sigma} = 2 \pi \sum_k | u^\alpha_{kj \sigma}|^2$
\cite{phasecancellations}.
Neglecting the $\alpha j\sigma$ dependence of $\gamma$,
the current thus is governed by 
the conduction electron DOS, $A_c (\omega ) = 
\sum_{j \sigma} A_{c, j\sigma}(\omega )$. Exploiting a 
Dyson equation for $G_{c, j\sigma}$, it has the form
\begin{eqnarray}
\nonumber
A_c (\omega) =  - {1 \over \pi}\sum_{j \sigma} 
\mbox{Im} \left[G_{0,j\sigma} (\omega) + v^2 
[G_{c, j\sigma}^{(0)} (\omega)]^2 G_{d\sigma}(\omega) \right] ,
\end{eqnarray}
where $G^{(0)}_{c, j\sigma} = 1/(\omega - \varepsilon_j + i 0^+)$ is the
free conduction electron  Green's  function \cite{broadening},
and the corresponding Kondo contribution to the
conductance  $G(V) = dI(V)/dV = G_0 (V) + \delta G(V)$
is \begin{eqnarray}
\delta G(V) = - \frac{e^2}{\hbar} &\gamma & \frac{\Gamma}{\pi}
\Delta \sum _{j,\sigma} {\cal P}\hspace {-1em} \int d\omega  
             A_{d\sigma } (\omega )  \times 
\label{conductance} \label{deltaG}
\\ 
&\biggl[ &\frac{\tilde F_V(\omega )- \tilde F_V(\varepsilon _j )}
            {(\omega - \varepsilon _j)^2} -
       \frac{d\tilde F_V(\omega )/d\omega }{\omega  -\varepsilon _j}
\biggr],  \nonumber \vspace{-1mm}
\end{eqnarray}
with $\tilde F_V(\omega ) = -\frac{d}{d\omega}(f(\omega -eV/2)+f(\omega
+eV/2))/2$. Even though Kondo physics appears only in the subleading 
contributions to
$A_c (\omega)$ and $G(V)$,  these are 
proportional 
\begin{figure}
\centerline{\psfig{figure=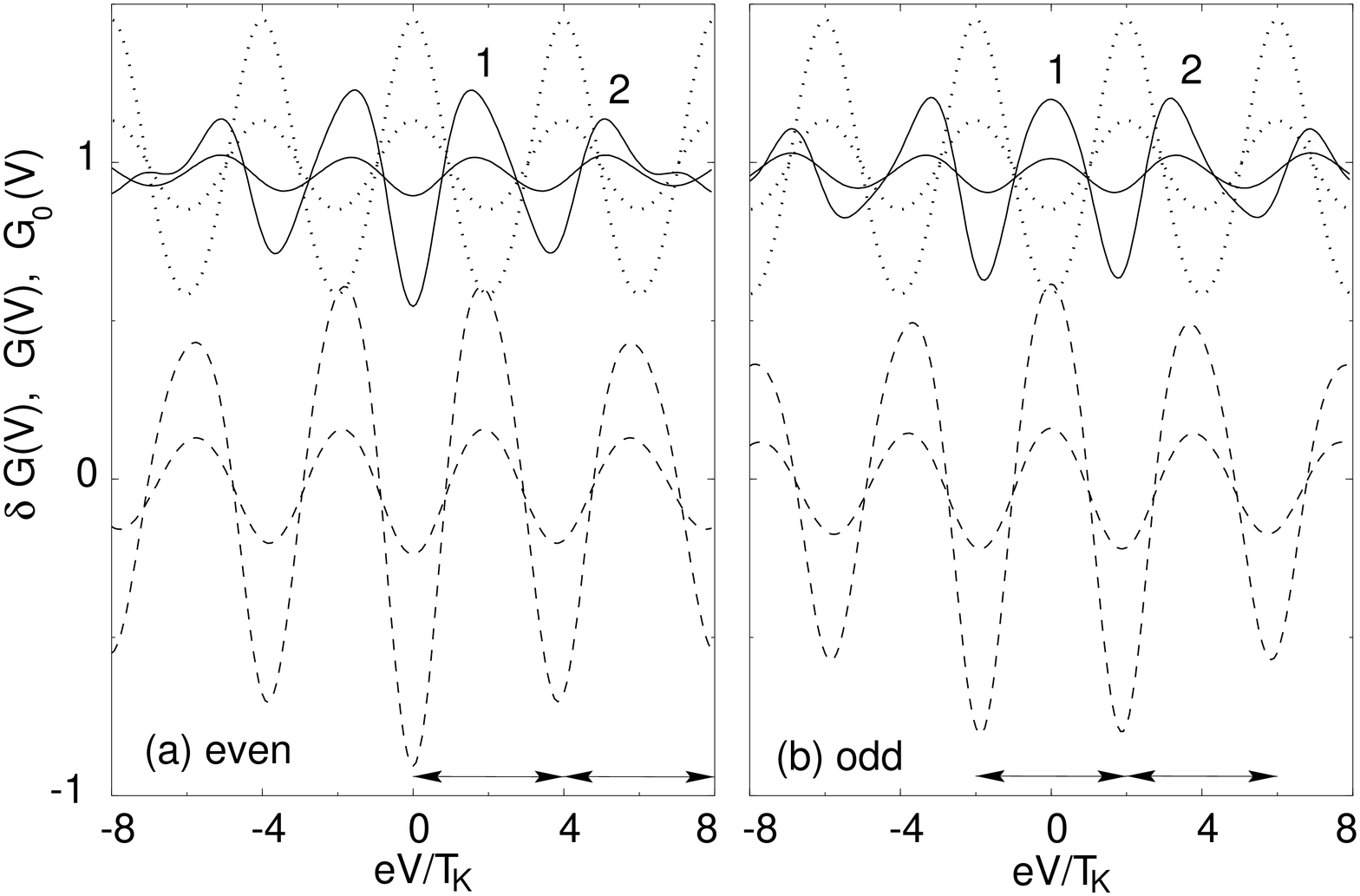,width=8.5cm}}
\narrowtext
\caption{Differential conductance of (a) an even and (b)
and odd   Kondo box for $\Delta= 4 T_K$. 
 Dotted, dashed and solid lines give, respectively,
the bare conductance $G_0(V)$, the 
Kondo contribution $\delta G(V)$, and the total
conductance $G_0 + \delta G$. The curves with larger or
smaller amplitudes correspond to $T = 0.7T_K$ and
$1T_K$, respectively; all are normalized such that the
average of the bare  conductance $\overline{G_0 (V)}= 1$.
}
\label{cond}
\end{figure}\noindent
\noindent
 to $v^2 = \Gamma \Delta/\pi$ and thus {\em grow}\/ 
with decreasing grain size.

$\delta G(V)$ and $G(V)$ are  shown in Fig.~\ref{cond}
and have rather irregular structures and line-shapes.  The reason for
this lies in the {\em interference}\/ between
$G_{d \sigma}$ and $[G^{(0)}_{c,j\sigma}]^2$ in $A_{c}$,
and correspondingly between $A_{d\sigma}$ and 
the bracketed factor in (\ref{deltaG}) for $\delta G(V)$. This
interference is reminiscent of a Fano resonance \cite{fano.61},
which likewise arises from the interference
between a resonance and the conduction electron DOS.  
Incidentally, Fano-like interference  
has been observed in STM spectroscopy of a single Kondo ion on a
metal surface \cite{crommie.98}, for which
the conduction electron DOS is flat. In contrast, for an ultrasmall grain
it consists of discrete peaks,
reflected in the last factor  in Eq. (\ref{deltaG}). This
leads to a much more complex interference pattern,
which does not directly mirror the specific peak structure 
 of $A_{d\sigma} (\omega)$ discussed above.

\begin{figure}
\centerline{\psfig{figure=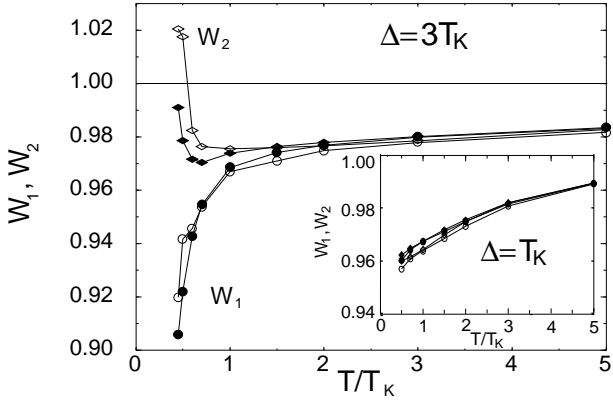,width=8cm}}
\narrowtext
\caption{Anomalous temperature
dependence  of the weights $W_1$ (circles) and $W_2$ (diamonds) of
the first two conductance peaks of
an even grain ($\circ$,$\lozenge$) and
an odd grain  ($\bullet$,$\blacklozenge$),  for $\Delta = 3 T_K$
and $\Delta = T_K$ (inset). We calculated the weights
under $G(V)$ over a fixed range $V = \Delta$
between successive minima of the lowest-$T$ $\delta G (V)$ curve
(see arrows in Fig.\ 3). 
}
\label{cweight}
\end{figure}

Nevertheless, $G(V)$ does bear observable traces of the Kondo effect,
in that {\em the interference pattern shows a distinct,
anomalous $T$-dependence,}\/ due to that of the Kondo resonance.
In particular, the weights $W_j$ under the individual peaks of $G(V)$ 
become $T$-dependent. (In contrast, the weight $W_0$ under
an individual peak of the bare conductance $G_0(V)$ is $T$ independent,
since the $T$ dependence of the peak shapes of $G_0$ 
are determined solely by $d \, f(\omega) / d \omega$.)      
This is illustrated in Fig.~\ref{cweight}, which  shows 
the $T$ dependence of the
weights $W_1$ and $W_2$ of the first and second conductance peaks
(counted relative to $V=0$ and labelled 1,2 in Fig.~\ref{cond}):
When $T$ decreases at fixed  $\Delta = T_K$, both $W_1$ and 
$W_2$ decrease, while at fixed $\Delta= 3 T_K$, $W_1$ decreases 
whereas $W_2$  increases.
The fact that the weights 
can either increase  or decrease with decreasing $T$
results from the constructive or destructive
 Fano-like interference effects discussed above.
Moreover, at the larger value for $\Delta$,
both $W_1$ and $W_2$ develop  a 
\noindent
{\em   parity effect
in the strength of their $T$ dependence.}\/ 
Since 
the peak weights in Fig.~\ref{cweight}
change by up to $\sim 10 \%$ as the grain is cooled
below $T_K$, it should be possible to 
experimentally \cite{ralph.97} detect 
their Kondo-induced anomalous 
$T$-dependence.
 
{\em Coherence length:---}\/
The condition $\Delta > T_K$ 
implies a relation between sample volume and 
the much-discussed spin coherence length $\xi_K = 2 \pi v_F / T_K$,
namely  (in 3D) ${\em Vol}\/ < \pi \xi_K / k_F^2$.
 Note that this relation
involves both the small length scale $1/k_F$
and  the sample's {\em volume,}\/ 
and not the smallest of  its linear dimensions, say $L$.
This implies  that  the length scale 
below which purely finite-size induced modifications
of  Kondo physics can  be expected is {\em not}\/ set by  $\xi_K$
alone \cite{Bergmann}, and indeed may be considerably smaller than $\xi _K$. 
This is why such modifications
were not found in the numerous  
recent experiments having $L \lesssim \xi _K$  for one or two sample
dimensions \cite{giordano,zawadowski}.

In conclusion, we have analysed the Kondo effect in an ultrasmall metallic
grain containing a single magnetic ion. 
The presence of a new energy scale in the system, the mean level
spacing $\Delta$,
leads to rich physical beha\-vior when $\Delta \sim T_K$, 
including a distinct even/odd effect. 
Our NCA calculations, which  
give a semi-quantitatively reliable estimate
of the size of the effects to be expected in 
future experiments, predict that the SET conductance through such a grain
has  a Kondo-induced anomalous $T$-dependendence of up to 10\%. 
Since the  effects discussed in this work
result, above all, from 
the discrete density of states near
$\varepsilon_F$, 
they should be generic for ultrasmall grains, 
i.e.\ robust against including randomness in the model, like 
$j$-dependent level spacings $\Delta_j$ and hybridization matrix 
elements $v_j$.
An interesting extension 
of this work would be to study  the effects of a magnetic field, which
couples not only to the magnetic impurity but also
Zeeman-splits each discrete $j$-level on the grain.

We are grateful to V. Ambegaokar, S. B\"ocker, F.Braun,
T. Costi, C. Dobler, P. Fulde,  A. Garg, D. Ralph, A. Rosch, G. Sch\"on, 
N. Wingreen, P. W\"olfle and G. Zar\'and for useful
discussions. This work was
started at an ISI Euroconference in Torino, and was supported by 
the DFG through  \vspace{-0.7cm}SFB195.

%%%%%%%%%%%%%%%%%%%%%%%%%%%%%%%%%%%%%%%%%%%%%%%%%%%%%%%
%%%%                  REFERENCES
%%%%%%%%%%%%%%%%%%%%%%%%%%%%%%%%%%%%%%%%%%%%%%%%%%%%%%%

%%%%%%%%%%%%%%%%%%%%%%%%%%%%%%%%%%%%%%%%%%%%%%%%%%
%%%%%%           FIGURE CAPTIONS
%%%%%%%%%%%%%%%%%%%%%%%%%%%%%%%%%%%%%%%%%%%%%%%%%%
\newpage
%%%%%%%%%%%%%%%
%%%%%%%%%%%%%%%
\end{multicols} 
%%%%%%%%%%%%%%%
%%%%%%%%%%%%%%%


\begin{references}
\bibitem{phasecancellations}
We neglect non-diagonal contributions to
$\gamma^\alpha$ of the form $2 \pi \sum_k
u^\alpha_{kj\sigma} u^{\alpha \ast}_{k j'\sigma'}$,
owing to the cancellations resulting from the random phases 
of the matrix elements.
\bibitem{broadening} 
When coupled to leads, the conduction 
band levels of the grain acquire a width $\simeq \gamma$.
We assume this to be negligible relative to $\Delta$, $T$  and $T_K$,
since realistic values \cite{ralph.97} yield $\gamma \lesssim 0.05 K$.
\end{references}
\end{document}